\documentclass[aps,pra,twocolumn,groupedaddress,showpacs]{revtex4}

\usepackage{graphicx}
\usepackage{graphicx}
\usepackage{amssymb}
\usepackage{amsmath}
\usepackage{color}
\newcommand{\be}{\begin{equation}}
\newcommand{\ee}{\end{equation}}
 
 \definecolor{BrickRed}{cmyk}{0,0.89,0.94,0.28}
\definecolor{MidnightBlue}{cmyk}{0.98,0.13,0,0.43}
\definecolor{DarkGreen}{rgb}{0,0.7,0.1}

\begin{document}

\title{Apparatus to probe the influence on the Casimir effect \\of the Mott-Anderson metal-insulator transition in doped semiconductors.}


\author{Giuseppe Bimonte}

\affiliation{ Dipartimento di Fisica E. Pancini, Universit\`{a} di
Napoli Federico II, Complesso Universitario
di Monte S. Angelo,  Via Cintia, I-80126 Napoli, Italy}
\affiliation{INFN Sezione di Napoli, I-80126 Napoli, Italy}

\email{giuseppe.bimonte@na.infn.it}

\begin{abstract}

We describe an isoelectronic differential apparatus  designed to observe the influence on the Casimir force  of the Mott-Anderson metal-insulator transition in doped semiconductors. Alternative theories of dispersion forces lead to different predictions for this effect. The investigation of this problem by standard apparatus, based on absolute  measurements  of the Casimir force, is very difficult because  the effect  is  small in the region of sub-micron separations, where the Casimir force can be measured precisely. The differential apparatus described here is immune by design to several sources of error that blur the interpretation of Casimir experiments, like electrostatic patches, inaccurate determination of plates separation, surface roughness and errors in the optical data.  With the help of the proposed setup it should be possible to   establish conclusively which among the alternative theories of the Casimir effect for semiconducting test bodies is correct.

\end{abstract}


\maketitle

\section{Introduction}
\label{sec:intro}

The Casimir effect \cite{Casimir48} is the force between two polarizable (discharged) bodies, caused by quantum and thermal fluctuations of the electromagnetic field in the region of space bounded by the two bodies. Even if it was predicted long ago, the Casimir effect attracted widespread attention only during the last decades, because the availability of new experimental techniques for the observation of small forces acting between macroscopic bodies at submicron separations, made it possible for the first time to accurately measure the tiny Casimir force, and study in detail its properties. For a review of the diverse roles played by the Casimir effect in both fundamental and applied physics, we address the reader to the books \cite{book1,parse,book2,RMP} and to the review articles \cite{capasso,woods}.

In his pioneering work \cite{Casimir48} Casimir calculated the force  between two ideal plane-parallel plates at zero temperature. The investigation of the Casimir effect in real material media started with the  fundamental paper of Lifshitz \cite{lifs},  which presented a derivation of  the force between two plane-parallel dielectric slabs in vacuum, at finite temperature. In his work, Lifshitz made use of the then new  theory of    
electromagnetic fluctuations developed by Rytov \cite{rytov}. Nowadays Rytov's theory has blossomed to a vast field of research, with many diverse applications extending from heat radiation to heat transfer, as well as to  Casimir and van der Waals forces in non-equilibrium situations etc. For a recent overview of fluctuational electrodynamics, as this field is called today, the reader may consult the article \cite{mehran}. 

Lifshitz theory of the Casimir effect is based  on the calculation of the stress-tensor for the fluctuations of the electromagnetic field in the vacuum region between the two bodies. By making use of the fluctuation-dissipation theorem, the latter fluctuations can be expressed in terms of the macroscopic electromagnetic response functions characterizing the bodies,  i.e. their respective electric (and magnetic) permittivities at imaginary values of the frequency $\omega$. Since the time of Lifshitz, the theory of the Casimir effect has been extended to arbitrary geometries of the bodies, by using modern scattering methods (see \cite{mehran} and references therein). The general theory is formulated in terms of the Matsubara Green's function of the electromagnetic field,     which depends on the  $T$-matrices of the two dielectric bodies, computed for imaginary frequencies. 

The numerous experiments carried over during the last 20 years \cite{book2,capasso,woods,mehran}, which used dielectric bodies of diverse materials and shapes, have shown in general good agreement with theoretical predictions based on Lifshitz theory,  within a few percent for submicron separations of the test bodies.  This is a  remarkable achievement indeed, in view of the macroscopic character of the Casimir effect, testifying to great efforts made both by experimentalists and theoreticians over the years.  To reach this goal, the surfaces of the bodies have to be fabricated with great care, their separation has to be determined with nanometer precision, and several potential sources of error need to be carefully scrutinized, like for example  the influence of surface curvature \cite{book2} and roughness \cite{book2,George}, possible issues with electrostatic calibrations \cite{Kim,Iannuzzi} and/or electrostatic patches on the surfaces of the bodies \cite{speake,behunin}.  Among  the potential sources of systematic errors,   the  importance of an accurate determination of the electric permittivity of the involved materials deserves a special mention. The crucial importance of this quantity for an accurate prediction of the Casimir force,  especially in the commonly adopted case of metallic bodies, is now widely recognized \cite{vitaly}. Nowadays, it is a common practice  to  measure the optical properties of the actual bodies that are used in the experiment. This data are then used  to compute, via dispersion relations \cite{book2}, the electric permittivity for the experimentally inaccessible imaginary frequencies entering Lifshitz formula. Special forms of dispersion relations have been devised \cite{bimonteK,bimonteKK}, to reduce the impact of an incomplete knowledge of the optical data on the low-frequency side, an issue of special importance in the case of conductors.

The  optimistic scenario outlined  above should not make one think that everything is settled and in good order in Casimir physics, because two series of  precise experiments with metallic bodies carried out by two distinct groups, one using micro-mechanical oscillators \cite{decca1,decca2,decca3,decca4} and the other atomic force microscopes  \cite{chang,bani1,bani2}  have revealed small, but nevertheless significant  deviations from theoretical predictions based on Lifshitz theory. As a premise to the discussion of these experiments, we recall that according to our current understanding of Lifshitz theory, based on the fluctuation-dissipation theorem, the response functions of the bodies to be used in the computation of the Casimir force should coincide with those that describe  their response to real external electromagnetic fields, as can be measured in an optical experiment. Surprisingly, it  appears that this expectation is contradicted by the precise experiments listed above.  The results of these experiments  have been shown to be inconsistent with Lifshitz theory, if in the computation of the Casimir force  the metallic surfaces are modeled at low frequency by the  familiar (lossy) Drude model, which is known to provide the correct description of ohmic conductors for low frequency. It appears instead that agreement with data is restored if the  conductors are modeled as lossless plasmas!  Agreement with the Drude model has been reported in a single torsion-balance experiment \cite{lamor}, probing the Casimir force up to the large separation of 7.3 $\mu$m.  The  interpretation of this latter experiment is, however, partly obscured by the fact that the Casimir force  was not measured directly, but rather  estimated indirectly after subtracting from the data  a much larger force, supposedly originating from electrostatic patches, by a fit procedure based on a phenomenological model of the unknown electrostatic force.

Motivated by these unexpected findings, some researchers felt the need for new theoretical criteria to select the low-frequency prescription for the material response functions, to be used in Lifshitz theory.   
A viable criterion that has been found is consistency of the chosen prescription with Nernst heat theorem.
Detailed analysis  \cite{bez1,bez2,bez3,bordag} has in fact demonstrated that the low frequency behavior of the Drude permittivity leads to violation of the Nernst theorem, in the idealized limit of conductors with perfect crystal lattices, while the plasma prescription is in agreement with that theorem. The  picture provided by the general principles of statistical mechanics is not  totally unequivocal though,  because later studies have shown that the Drude model  is consistent with the
Bohr-van~Leeuwen theorem of classical statistical physics,  while the plasma model is not \cite{bimontebohr}.

The attitude of the community towards the above conundrum is mixed. The opinion has been expressed that the experimental evidence in favor of the plasma  prescription  is not really waterproof. After all, the observed discrepancies  which motivated this prescription are very small in the submicron separation region where the Casimir force can be measure accurately, something like one or two percent. It has been pointed out that  perhaps the observed small discrepancies are  due to small experimental errors that have escaped detection, like for example  small errors in the determination of the  bodies separation, or in the electrostatic calibration, the presence of patches on the surfaces, roughness or incomplete and/or inaccurate optical data. There is the widespread opinion that before abandoning the guidance of principles as fundamental as the fluctuation-dissipation theorem, one should be  sure that there is a crisis.

More puzzles  connected with the influence of free charge carriers  on   the Casimir force, have emerged from  recent experiments  \cite{umar2006, umar2006bis,umar1} with semiconducting plates.  Investigations of the Casimir effect with semiconductors  are of special interest, in view of the unique role played by 
these materials in modern technology, where they represent the  reference materials  for the fabrication of optomechanical, micro- and nano-mechanical devices. The process of miniaturization of these machines has now reached the point  where the Casimir interaction between their movable constituent parts is often comparable with electrostatic forces. In such circumstances, the Casimir force can either appear as a nuisance capable of perturbing the correct operation of the machine, possibly determining stiction and adhesion of its parts \cite{Buks},  or rather as a useful resource  that can be harnessed  to operate the device in new ways \cite{capasso2}.  This explains why  the study of the  Casimir effect between semiconducting bodies has been pursued intensely in recent years.  

Apart from technology, the unique properties of semiconductors  make them  a very interesting tool to investigate aspects of Casimir physics relating to relaxation phenomena in  conductors.  As it is well known \cite{ash}, intrinsic semiconductors are insulators at zero temperature, and even at room temperature their resistivity is very high,  due to the low density of thermally excited free charge carriers. The conductivity of semiconductors can however be greatly increased   by doping, and it has been known for a long time \cite{rosen} that for concentrations of dopants higher than a critical density $n_{\rm cr}$ (which depends on both the semiconductor and the dopant) doped semiconductors undergo a Mott-Anderson insulator-metal transition. The possibility of altering by  doping the conductivity of semiconductors by several orders of magnitudes,   prompted some researches  to investigate whether  the Casimir force can be modified by changing the carrier density of a semiconductor plate. That this is indeed possible was  demonstrated in \cite{umar2006}, where the Casimir forces between a gold coated sphere and two P-doped Si plates of different carrier densities were measured and compared.  The carrier densities  of the two Si plates,  $n_a = 1.2 \times 10^{16} {\rm cm}^{-3}$ and  $n_b = 3.2 \times 10^{20} {\rm cm}^{-3}$, were chosen to be  respectively lower and larger than  the Si critical density $n_{\rm cr}=3.84 \times 10^{18} {\rm cm}^{-3}$. In accordance with Lifshitz theory, the Casimir  force was found to have a larger magnitude for the plate of lower resistivity.  More precisely, the measured force for the plate of higher resistivity  is in agreement with the  value obtained by plugging the tabulated optical data \cite{Palik} of intrinsic Si into Lifshitz formula, while the larger force observed for  the plate of lower resistivity can be reproduced by Lifshitz formula by augmenting the permittivity of intrinsic Si by a Drude contribution accounting for the density of free carriers.  A successive more precise experiment \cite{umar2006bis} with a B-doped Si plate having a carrier density $n \approx 3 \times 10^{19} {\rm cm}^{-3}$ allowed to establish that the measured Casimir force, while consistent with inclusion of the contribution of free carriers in the electric permittivity of the plate, is in fact inconsistent with neglect of such a contribution. In another experiment \cite{umar1} it was demonstrated that the Casimir force between a gold coated sphere and a Si membrane can be modified by  laser illumination. The observed change in the Casimir force results from the large  increase in the carrier density of the Si membrane caused by laser illumination, from its room-temperature equilibrium value ${\tilde n}= 5 \times 10^{14} \,{\rm cm}^{-3}$ to a value larger than $10^{19} \,{\rm cm}^{-3}$ in the presence of light. The analysis of the data shows that the observed change in the Casimir force is consistent  with the theoretical prediction, if and only if the contribution of free carriers is included in Lifshitz formula when computing the force in the presence of light, and fully excluded from it when computing the force in the absence of light. 

A very interesting question that can be asked about the Casimir effect in semiconductors is whether the metal-insulator transition has any influence on the Casimir force at room temperature. The theoretical answer to this question depends crucially on the prescription that is used to describe in Lifshitz formula the influence of free carriers in doped semiconductors. According to the standard prescription, based on the fluctuation-dissipation theorem, no effect is to be expected since the optical properties of semiconductors at room temperature  do not change appreciably across the transition.  A  different  theoretical approach leads however to the bold prediction that the Casimir force should display a {\it discontinuous}   change  across the transition!  Let us see how this comes about. 

To assess  whether the contribution of the free carriers should be included or excluded from the determination of the Casimir force, recourse can  again be made to the criterion of consistency with the Nernst heat theorem.  One then finds \cite{ther1,ther2,ther3,ther4} that in materials that are  insulators at zero temperature, the theorem is violated if the temperature-dependent contribution of thermally excited carriers is included in the permittivity. These findings led the authors of \cite{vladimir} to the following  prescription for semiconductors: free charge carriers of doped semiconductors   {\it do contribute} to the Casimir force if and only if the semiconductor is in the metallic phase, i.e. for charge concentrations {\it larger} than the critical density $n_{\rm cr}$ for the insulator-metal transition. Instead, the contribution of charge carriers has to be {\it neglected} when the semiconductor is in the insulator state, i.e. for densities $n$ {\it smaller} than $n_{\rm cr}$. So, similarly to  metals, we have here another instance of a   prescription for the low-frequency response of a semiconductor, which is in sharp contrast with the observed response of these materials  to real external electromagnetic fields.  We point out that there is no consensus on the thermodynamic  argument that led the authors of \cite{vladimir} to formulate this prescription.  The claimed violation of Nernst heat theorem by insulators with dc conductivity included,   was  in fact proven  on the basis of the  standard  formulation  of Lifshitz theory, in which the material is characterized by a {\it local} response function. The validity of this approach   for conductors with a small  density of carriers  has been criticized by other investigators \cite{pita,vitaly2,diego2}, on the basis of the observation that the response function of poor conductors at low frequency is {\it non-local}, because of incomplete screening of electric fields (Debye screening). It has been shown however \cite{vladpita,vladdiego} that the non-local approach of  \cite{pita}  leads to predictions for the Casimir force that are in disagreements with the precise measurements of the experiment \cite{umar1}.  

A striking implication of the above prescription is that  the Casimir force among semiconducting test bodies should display a {\it discontinuous} change, as the carriers density of the semiconductor  traverses  the critical value $n_{\rm cr}$. What is striking here is that  the discontinuous change in the Casimir force  occurs without a detectable change in the  optical properties of the semiconductor!  The authors of  \cite{vladimir} observe that the possibility of having a change, and in fact a very large change,  in the Casimir force in the absence of a detectable change of the optical properties of the plate, has been indeed demonstrated by an experiment with an indium tin oxide  (ITO) film  \cite{umar2,umar3}. In this experiment it was shown that the Casimir force between a Au coated sphere and an ITO film deposited on a quartz substrate  can be decreased up to 35 \% by UV treatment of the ITO film. Ellipsometry measurements of the imaginary part of the permittivity of the untreated and UV treated ITO film showed no significant differences, which led the authors to conjecture that the observed change in the Casimir force was determined by a phase transition of the ITO film from a metallic to an insulator state, caused by the UV treatment. As a final remark about the prescription proposed in \cite{vladimir}, we would like to point out that  the authors did not explicitly address the question whether in the metallic phase the carriers contribution to the Casimir force should be described by the lossy Drude model, or rather by the lossless plasma model.  Analogy with ordinary metals  suggests that the plasma model provides the correct description. We note that the precision of the experiments with semiconductors quoted earlier is not sufficient to discriminate the Drude model from the plasma model, and so the question remains open.

The above considerations motivated us to see if it is possible to  observe experimentally the variation of the Casimir force across the metal-insulator transition predicted in \cite{vladimir}.  Achieving this goal by  measuring the absolute Casimir force with an ordinary apparatus is very difficult, because the effect predicted by the theory of \cite{vladimir} is small (a few percent) in the region of submicron separations where the force can be measured precisely. In this paper we demonstrate that an {\it isoelectronic differential} setup  may provide the answer (see Fig. \ref{setup}). The proposed setup would also allow to establish  whether the Drude or plasma prescriptions provide the correct description of the influence of the charge carriers on the Casimir force in the conducting state of the semiconductor.   
 
Isoelectronic differential Casimir setups were proposed by us a few years ago \cite{bimoiso1,bimoiso2,bimoiso3}  to help resolving the Drude-plasma conundrum with ohmic  conductors, whether non-magnetic or magnetic. It is well known that differential force measurements offer great advantages, compared to absolute force measurements,  since they have a much higher sensitivity.  Sensitivities  of one or two fN in difference-force measurements have been already reported in the literature \cite{ricardomag,umardiff}, which are a thousand times larger than the typical pN sensitivity of  modern absolute-force measurement apparatus.  Another advantage results from the fact that  the differential measurement  is performed by executing a  small lateral displacement (a few tens of microns) of  the sensing apparatus in a plane parallel  to a structured plate  (or viceversa as in the experiment \cite{ricardomag}). This  procedure leads to cancellation of   uncertainties in the vertical separation among the surfaces, which represent a delicate problem in absolute Casimir measurements. A differential setup  inspired to these principles was indeed proposed in \cite{umardiff} to  observe the difference among the Casimir forces between a Au coated sphere and the two sectors of a  structured Si surface, characterized by  different carrier densities. The nice configuration of \cite{umardiff}  goes into the right direction, but in our opinion it still presents a potential drawback, since  the exposed surfaces of the  two Si sectors  may have  {\it different}  potential patches as a result of their different dopings.  If this indeed happens,  a spurious differential force  of electrostatic origin among the two sectors of the plate arises, which could be very harmful in principle.  The resolution of this problem brings us to the second ingredient of our setup, i.e. the {\it isoelectronic} scheme, which consists in   covering   the structured plate used for the differential measurement with a thin homogeneous conductive layer. The over-layer provides an electrostatic screen, which neutralizes the effect of possible inhomogeneities on the surface of the structured plate. Eventual stray electrostatic forces that may be caused by patches on the exposed surface of the over-layer  are uniform with respect to the position of the sensing apparatus over the plate,  and therefore they automatically cancel out from the differential force (up to small statistical fluctuations \cite{behu3}). By the same token, the differential isoelectronic scheme ensures of course  cancellation of surface roughness effects   \cite{bimoiso1,bimoiso2,bimoiso3}. Isoelectronic setups   were pioneered by the Purdue group in  Casimir-less experiments searching for Yukawa type corrections to Newtonian gravity \cite{deccaiso1,deccaiso2}. The power of differential isoelectronic setups in Casimir experiments has been demonstrated by a recent experiment \cite{ricardomag} which measured the force-difference between a Au or Ni coated sphere and alternating Ni-Au sectors of a micro-fabricated rotating disk, covered by a thin Au over-layer.  In this experiment the isoelectronic configuration led to a thousandfold  amplification of the difference among the Drude and plasma prescriptions for magnetic materials \cite{bani1,bani2}, which allowed for an unambiguous discrimination among them. In particular, the Drude model with inclusion of the magnetic properties of Ni was unequivocally ruled out, while the plasma model with inclusion of the Ni magnetic properties was found to be in good agreement with the data. The experiment showed also that neither the Drude nor the plasma model with exclusion of the Ni magnetic  properties could account for the observations. Very recently, we also proposed an isoelectronic setup to probe the influence on the Casimir force of relaxation phenomena in metals and in doped semiconductors in the dielectric state \cite{bimoiso4}. 

The plan of the paper is as follows: in Sec. II we describe our differential apparatus, and present the general formalism for the computation of the differential Casimir force. In Sec. III we discuss alternative prescriptions that have been proposed in the literature to deal with the influence of free charge carriers on the Casimir force between conducting and semiconducting  test bodies. In Sec IV we present   our numerical computations of the differential force in our apparatus, and discuss the impact of several possible systematic errors. Finally, in Sec V we present our conclusions.

\section{Isoelectronic setup and general formalism}

\begin{figure}
\includegraphics [width=.9\columnwidth]{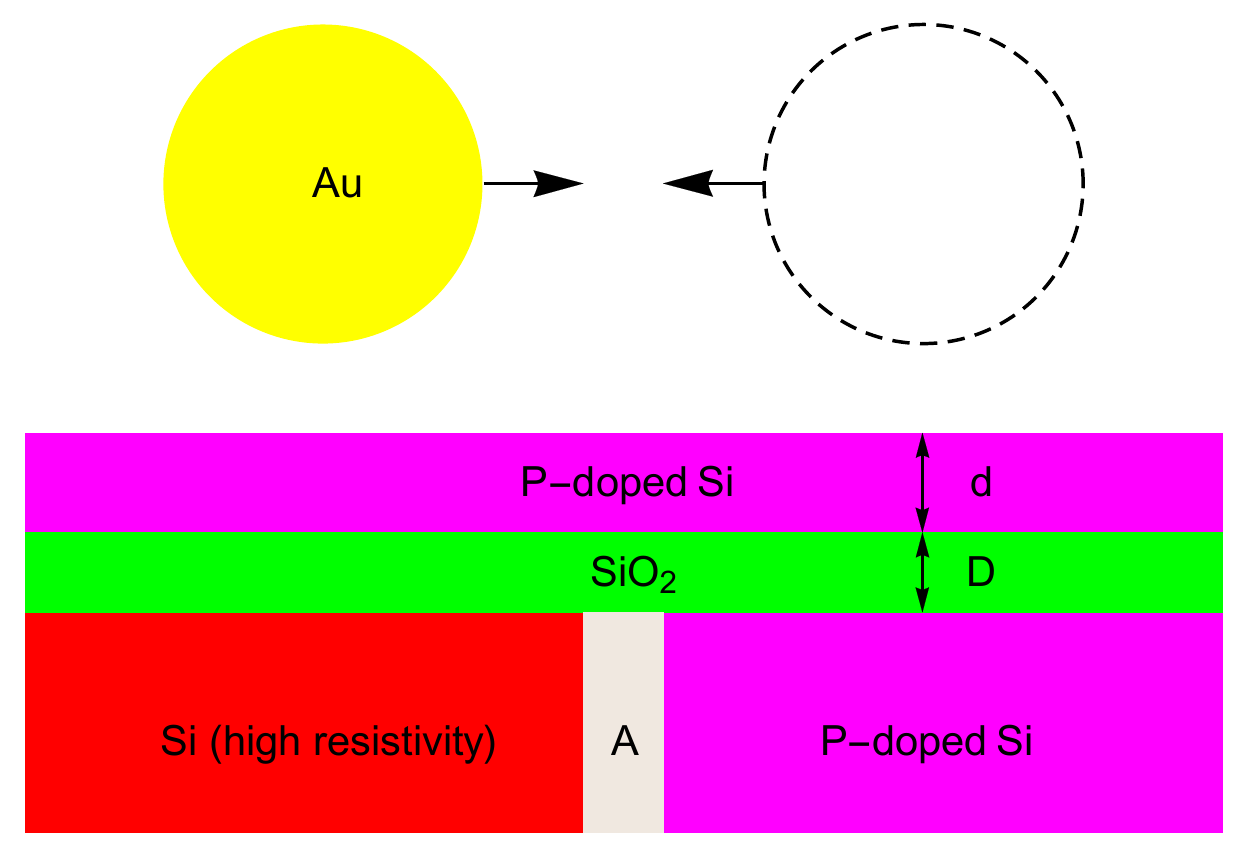}
\caption{\label{setup}  Isoelectronic differential setup: a Au-coated sphere can be moved in a plane parallel to a micro-fabricated patterned Si plate, consisting of two differently doped regions. The measured quantity is the differential Casimir force between the sphere and the two regions of the Si plate. The isoelectronic configuration is realized by covering the plate  with a uniform conductive Si over-layer of thickness $d=50$ nm.  To prevent carrier diffusion from the over-layer to the undoped left sector of the plate, a thin ${\rm SiO}_2$  inter-layer of thickness $D=10$ nm  is interposed between the over-layer and the bottom Si sectors.  }
\end{figure}

We consider the configuration of a Au-coated sphere with radius $R=150\;\mu$m in vacuum, at a (minimum) distance $a$ from a microfabricated patterned plate at room temperature $T=300$ K. The thickness of the Au coating of the sphere is supposed to be larger than 100 nm, which allows 
to consider it as if it were made entirely of Au in our computations of the Casimir force. The key ingredient of our setup is the micro-fabricated plate. Its  structure  is illustrated in Fig. \ref{setup}: its right half is made of P-doped Si, while its left half is made of high resistivity Si.  The thickness of both sectors is supposed to be large enough to consider both as infinitely thick in the computation of the Casimir force. In order to realize an isoelectronic configuration, both sectors are covered with a conductive over-layer. As it was explained in the Introduction, the purpose of the over-layer  is to screen out potentially detrimental electrostatic forces caused by non-uniform potential patches on the surfaces of the differently doped Si sectors of the plate.  At the same time, the over-layer should be semi-transparent, in order for the Au sphere to be able to  ``see'' the underlying Si sectors of the plate.  These two demands can be met by choosing for the over-layer a material  whose conductivity is large enough  to ensure screening of electrostatic fields, but not so large to make it opaque. The latter constraint leads one to exclude Au, since its small plasma length $\lambda_p$  renders a  Au over-layer exceedingly opaque for our purposes.  These considerations led us to consider  P-doped Si  as  a possible material for the over-layer.  We recall that P-doped conductive  Si plates have been already successfully utilized  in precision Casimir experiments \cite{umar2006}.  In our setup we consider an over-layer with a thickness $d=50$ nm.
For simplicity, we have assumed in our computations that the carrier density $n$ of the over-layer is the same as that of the right sector, but this assumption is by no means necessary. To prevent carrier diffusion from the over-layer to the undoped left sector of the plate, a thin insulating ${\rm SiO}_2$ layer of thickness $D=10$ nm  is interposed between the over-layer and the bottom Si sectors. The light-colored sector (marked by the letter A) separating the left and right sectors of the plate has again the purpose of preventing carrier diffusion among the two sectors. Its material needs not be specified for our purposes. 
We imagine that the sphere can be moved in a plane parallel to the surface of the patterned plate, from a position $P_{\rm ins}$ (represented by the filled yellow disk in Fig. \ref{setup})    to the position $P_{\rm dop}$ (represented by the dashed empty disk in Fig. \ref{setup}). We shall denote as $F_{\rm ins}(a,T)$ and  $F_{\rm dop}(a,T)$ the respective Casimir forces on the sphere.  It is assumed that the vertical projections of the points $P_{\rm ins}$ and $P_{\rm dop}$ lie deep into the left and right sectors of the plate, respectively \cite{bimoiso3}. This ensures that in the computation of the forces $F_{\rm ins}(a,T)$ and  $F_{\rm dop}(a,T)$  one can neglect the sharp boundary separating the left and the right halves of the plate, and treat both sectors as infinitely wide in all directions in the horizontal plane. The quantity of interest to us is the {\it differential} force $F_{\rm diff}(a,T)$:
\be
F_{\rm diff}(a,T)=F_{\rm ins}(a,T)-F_{\rm dop}(a,T)\;.
\ee 
For separations $a \ll R$, the force can be estimated using the proximity force approximation (PFA) \cite{book2}. It has been recently shown that the error implied by PFA   in the sphere-plate force is  smaller than $a/R$ \cite{jaffe,apl,teo,bimoprec,gert}. For our configuration, the PFA gives:
\begin{eqnarray}
&&
F_{\rm diff}(a,T)=k_B T R\,
\sum_{l=0}^{\infty}{\vphantom{\sum}}^{\prime}
\int_{0}^{\infty}\!\! k_{\perp} d k_{\perp}
\label{eq3} \\
&&\times \!\!
\sum_{\alpha }
\ln\frac{1-e^{-2 a q_l}\,r_{\alpha}^{(0,1)}(i\xi_l,k_{\perp})R_{\alpha}^{(0,2,3,4)}(i\xi_l,k_{\perp})
}{1-e^{-2 a q_l}\,r_{\alpha}^{(0,1)}(i\xi_l,k_{\perp})R_{\alpha}^{(0,2,3,2)}(i\xi_l,k_{\perp})},\nonumber \label{lifs}
\end{eqnarray}
where $k_B$ is Boltzmann constant, $\xi_l=2 \pi l k_B T/\hbar$ are the (imaginary) Matsubara frequencies,
$k_{\perp}$ is the modulus of the in-plane wave-vector, $q_l=\sqrt{\xi_l^2/c^2+k_{\perp}^2}$, and the prime in the summation sign indicates that the $l=0$ term is taken with a weight 1/2. The summation over $\alpha$ is taken over the two independent polarizations of the electromagnetic field, i.e. the transverse magnetic (TM) and the transverse electric (TE) modes. To explain the meanings of the reflection coefficients  $r_{\alpha}^{(0,1)}(i\xi_l,k_{\perp})$, $R_{\alpha}^{(0,2,3,2)}(i\xi_l,k_{\perp})$ and $R_{\alpha}^{(0,2,3,4)}(i\xi_l,k_{\perp})$  that occur in the above Equation, we introduce the following notations. The four materials that constitute our setup, i.e.  Au, P-doped Si, ${\rm SiO}_2$ and high-resistivity Si shall be distinguished by the labels  $p=1,2,3,4$ respectively, and their permittivities shall accordingly be denoted as $\epsilon^{(p)}(i \xi_l) \equiv \epsilon^{(p)}_l$. The label $p=0$  denotes the vacuum, and we  set $\epsilon^{(0)}_l  \equiv 1$. Thus  $r_{\alpha}^{(p,q)}(i\xi_l,k_{\perp})$ denote the Fresnel reflection coefficients for a planar interface between media $p$ and $q$:
\begin{eqnarray}  
r_{\rm TE}^{(p,q)}(i \xi_l,k_{\perp}) &=& \frac{k_l^{(p)}-k_l^{(q)}}{k_l^{(p)}+k_l^{(q)}}\;,\\
r_{\rm TM}^{(p,q)}(i \xi_l,k_{\perp}) &=&  \frac{  \epsilon^{(q)}_l k_l^{(p)}-  \epsilon^{(p)}_l k_l^{(q)}}{ \epsilon^{(q)}_l k_l^{(p)}+  \epsilon^{(p)}_l k_l^{(q)}}\;,\label{fresnel}
\end{eqnarray}  
where $ k_l^{(p)}= \sqrt{ \epsilon^{(p)}_l \xi_l^2/c^2+k_{\perp}^2}$.
The symbols $R_{\alpha}^{(0,p,q,r)}(i\xi_l,k_{\perp})$ denote instead the reflection coefficients of a plane-parallel three-layer slab consisting of a thick slab of material $r$ covered by two layers made of the materials $p$ and $q$, of respective thicknesses $d$ and $D$, where $p$ is the material of the outer layer. The expression of  $R_{\alpha}^{(0,p,q,r)}(i\xi_l,k_{\perp})$ is:  
\be  
R_{\alpha}^{(0pqr)}({\rm i} \xi_l,k_{\perp})=\frac{r_{\alpha}^{(0 p)}+e^{-2\, d k_l^{(p)}}\,r_{\alpha}^{(pqr)}}{1+e^{-2\,d\, k_l^{(p)}}\,r_{\alpha}^{(0p)}\,r_{\alpha}^{(pqr)}}\;,\label{three1}
\ee   
where
\be  
r_{\alpha}^{(pqr)} =\frac{r_{\alpha}^{(pq)}+e^{-2\,D\, k_l^{(q)}}\,r_{\alpha}^{(qr)}}{1+e^{-2\,D\, k_l^{(q)}}\,r_{\alpha}^{(pq)}\,r_{\alpha}^{(qr)}}\;.\label{three2}
\ee  

\section{Three prescriptions for the Casimir force}

The Equations presented in the previous Section  can be used to compute the differential force,  after a specific prescription for the values of the permittivities $\epsilon^{(p)}_l$ is made. As it was explained in the Introduction, there exist in the literature three distinct prescriptions for computing the Casimir force between  test bodies made of metals and/or conductive doped semiconductors, which for brevity we shall refer to as  Drude model,  plasma prescription and  insulator state prescription (ISP) respectively.  Below, we shall discuss what they imply for our setup.

\subsection{The Drude model}

The Drude model represents what we may think of as the “ orthodox” formulation of Lifshitz theory. This formulation, which is based on the fluctuation-dissipation theorem, instructs us to use  for  $\epsilon^{(p)}_l$ the values that correspond to the analytic continuation to the imaginary axis of the “true” complex permittivities $\epsilon^{(p)}(\omega)$ of the materials, as can be measured in an optical measurement. As it is well known \cite{book2} knowledge of the imaginary part ${\rm Im}[\epsilon^{(p)}(\omega)]$ of the permittivity allows to determine $\epsilon^{(p)}_l$ via Kramers-Kronig relations, or their generalizations \cite{bimonteK,bimonteKK}. Ideally, in a concrete experiment  one would measure the optical data of the test bodies that constitute the apparatus.  In this work, we shall rely on the tabulated optical data for Au, Si and SiO$_2$  \cite{Palik} . 

In \cite{Palik} optical data for Au are listed  for frequencies larger than $0.125$ eV/$\hbar$. This is not sufficient to  compute  $\epsilon(i \xi_l)$ for small values of $l$ (since $\xi_1=0.16$ eV/$\hbar$).  Following the standard procedure \cite{book2}, the data  are extrapolated to low frequencies by  a Drude-like model of the form:
\be
\epsilon_{\rm Au}(\omega)=-\frac{\omega^2_1}{\omega(\omega + i \gamma_1)}+\epsilon^{\rm core}_{\rm Au}(\omega)\,,\label{gold}
\ee 
where $\epsilon^{\rm core}_{\rm Au}(\omega)$ accounts for the contribution of bound (core) electrons,  and $\omega_{1}$ and $\gamma_{1}$  are the plasma and the relaxation frequencies, respectively. We set   $\omega_{1}=9$ eV/$\hbar$ and $\gamma_{1}=0.035$ eV/$\hbar$ \cite{book2}. The  Drude model implies:
\be
\epsilon_l^{(1)}= \epsilon_{\rm Au}(i \xi_l)=\frac{\omega^2_1}{\xi_l(\xi_l + \gamma_1)}+\epsilon^{\rm core}_{\rm Au}(i \xi_l)\;.\label{goldDr}
\ee

The Si data of \cite{Palik} refer to intrinsic (highly resistive) Si, and  we shall denote the corresponding permittivity   by $\epsilon^{\rm int}_{\rm Si}(\omega)$.  For the highly-resistive Si constituting the left sector of our patterned plate  we thus set once and for all:
\be
\epsilon^{(4)}_l = \epsilon^{\rm int}_{\rm Si}(i \xi_l)\;.\label{insSi}
\ee

Now we consider P-doped Si. It is known \cite{Palik} that the permittivity of conductive Si  is well described by   the formula:
\be
\epsilon_{\rm Si}^{\rm cond}(\omega)=-\frac{\omega^2_2}{\omega(\omega + i \gamma_2)}+\epsilon^{\rm int}_{\rm Si}(\omega)\,,\label{sicon}
\ee
where the Drude term accounts for the contribution of free carriers.
The value of ${\omega_2}$ is related to the doping concentration $n$ by the formula:
\be
{\omega_2}=e \sqrt{\frac{4 \pi n}{m^*}}\;,
\ee
where  $e$ is the electron charge and $m^*$ is the effective electron mass (in P-doped Si the charge carriers are electrons). The value of the relaxation frequency $\gamma_2$  is related, via Eq. (\ref{sicon}), to the sample conductivity $\sigma_2$:
\be
\gamma_{2}=\frac{{\omega^2_2}}{4 \pi \sigma_2}\;.
\ee
We shall fix $\gamma_2= 5.5 \times 10^{13}$ rad/sec, which represents the value of the relaxation frequency  for a concentration $n=3.5 \times 10^{18} {\rm cm}^{-3}$ (close to the critical value for P-doped Si $n_{\rm cr}=3.84 \times 10^{18} {\rm cm}^{-3}$). The chosen value of $\gamma_2$ corresponds to the conductivity $\sigma_2 \approx 0.64 \times 10^{14} {\rm s}^{-1}$ \cite{silic}. We remark that the precise value of $\gamma_2$ is not  important for our purposes, because the force $F_{\rm diff}$ is only weakly dependent on $\gamma_2$. The  Drude model implies that for the conductive Si constituting both the over-layer and the right sector of our plate we should set:
\be
\epsilon_l^{(2)}=\epsilon_{\rm Si}^{\rm cond}(i \xi_l)=\frac{\omega^2_2}{\xi_l(\xi_l + \gamma_2)}+\epsilon^{\rm int}_{\rm Si}(i \xi_l)\;.
\ee
Finally, we consider SiO$_2$. This is an insulator. In our computations we fix once and for all:
\be
\epsilon^{(3)}_l= \epsilon_{\rm SiO_2}(i \xi_l)\;,\label{epsSiO2}
\ee
where for  $\epsilon_{\rm SiO_2}(\omega)$ we take the data in \cite{Palik}.  

 It is interesting to note that within the Drude model, the reflection coefficients $r_{\alpha}^{(0,1)}(i\xi_l,k_{\perp})$, $R_{\alpha}^{(0,2,3,2)}(i\xi_l,k_{\perp})$ and $R_{\alpha}^{(0,2,3,4)}(i\xi_l,k_{\perp})$ attain a {\it universal} value for vanishing frequency, i.e. for the Matsubara index $l=0$ that corresponds to the so-called classical term of Lifshitz formula. It is a simple matter to check that:
\be
 r_{\rm TM}^{(0,1)}(0,k_{\perp})=R_{\rm TM}^{(0,2,3,2)}(0,k_{\perp})=R_{\rm TM}^{(0,2,3,4)}(0,k_{\perp})=1\,,\label{univTM}
\ee
\be
 r_{\rm TE}^{(0,1)}(0,k_{\perp})=R_{\rm TE}^{(0,2,3,2)}(0,k_{\perp})=R_{\rm TE}^{(0,2,3,4)}(0,k_{\perp})=0\,.
\ee
According to Eq. (\ref{lifs})  this implies that the classical $l=0$ term contributes nothing to $F_{\rm diff}$ within the Drude  model.

\subsection{The plasma prescription}

As we discussed in the Introduction, thermodynamic considerations based on the Nernst theorem together with the results of several precise experiments motivated some investigators to propose a new prescription \cite{book2} for computing the Casimir force between metallic and/or semiconducting bodies.  This alternative prescription, that we shall refer to as the plasma prescription, posits the following rule \cite{book2,RMP}: \\

\noindent
{\it plasma prescription: if a  conductor is in the conducting state (i.e. is a conductor at $T=0$) the contribution of its free carriers  must be included in the computation of the Casimir force, but relaxation phenomena  must be neglected. In other words, its free carriers should be modeled at low frequency as a dissipation-less plasma.}\\

The implications of the above prescription for  the permittivities $\epsilon_l^{(p)}$ of the materials that constitute our setup are the following. The values $\epsilon_l^{(1)}$ of the Au permittivity should be computed using instead of Eq. (\ref{goldDr}) the following modified Equation:
\be
\epsilon_l^{(1)}\vert_{\rm pl}=\frac{\omega^2_1}{\xi_l^2}+ \epsilon^{\rm core}_{1}(i\xi_l)\;.
\ee 
Now we turn to conductive Si. As we explained in the Introduction, for carrier concentration $n>n_{\rm cr}$ a semiconductor is in the  conducting state (at $T=0$). Thus, the plasma prescription posits that for $n>n_{\rm cr}$ the permittivity $\epsilon_l^{(2)}$ of the conductive Si constituting the over-layer and the right sector of our patterned plate should be computed using the formula   
\be
\epsilon_l^{(2)}\vert_{\rm pl}=\frac{\omega^2_2}{\xi_l^2}+\epsilon^{\rm int}_{\rm Si}(i \xi_l)\;,\;\;\;{\rm for} \;n > n_{\rm cr}\;.
\ee
Of course, the permittivities $\epsilon_l^{(4)}$ of the highly resistive Si constituting the left sector of the plate, as well as the permittivity $\epsilon_l^{(3)}$ of the SiO$_2$ layer are still computed according to Eqs. (\ref{insSi}) and (\ref{epsSiO2}), respectively.
 It is interesting to note that   the plasma-model reflection coefficients are no more universal in the limit of vanishing frequency. More precisely, while for TM polarization the three zero-frequency reflection coefficients $r_{\rm TM}^{(0,1)}(0,k_{\perp})$, $R_{\rm TM}^{(0,2,3,2)}(0,k_{\perp})$ and $R_{\rm TM}^{(0,2,3,4)}(0,k_{\perp})$  remain equal to one,  as in the Drude  model (see Eq. (\ref{univTM})), the TE reflection coefficients have the following non-universal values:
\be
r_{\rm TE}^{(0,1)}(0,k_{\perp})\vert_{\rm pl}=\frac{k_{\perp}-s_1}{k_{\perp}+s_1}\;,
\ee 
\be
R_{\rm TE}^{(0,2,3,4)}(0,k_{\perp})\vert_{\rm pl}=  \frac{(k_{\perp}^2-s_2^2)\left(1-e^{-2 d  s_2} \right)}{(k_{\perp}+s_2)^2-e^{-2 d s_2} (k_{\perp}-s_2)^2}\;,
\ee
$$
R_{\rm TE}^{(0,2,3,2)}(0,k_{\perp})\vert_{\rm pl}=\frac{(k_{\perp}-s_2)}{(k_{\perp}+s_2)}
$$
\be
\times \; \frac{e^{2 d s_2}(k^2_{\perp}+s_2^2 + 2 k_{\perp} s_2 \coth(2 D k_{\perp}))-(k_{\perp}+s_2)^2}{e^{2 d s_2}(k^2_{\perp}+s_2^2 + 2 k_{\perp} s_2 \coth(2 D k_{\perp}))-(k_{\perp}-s_2)^2} \;,
\ee
where we set $s_p=\sqrt{\omega_p^2/c^2+ k_{\perp}^2}$, $p=1,2$. Notice that since $R_{\rm TE}^{(0,2,3,4)}(0,k_{\perp})\vert_{\rm pl} \neq R_{\rm TE}^{(0,2,3,2)}(0,k_{\perp})$, the classical $l=0$ Matsubara term for TE polarization does contribute to the differential force, within the plasma prescription. This fact, which marks an important difference between the plasma and the Drude prescriptions, constitutes the main reason of the different magnitudes predicted by the two prescriptions for the differential force $F_{\rm diff}$.

\subsection{The insulator state prescription} 

As we explained in the Introduction, thermodynamic considerations suggest to neglect the contribution to the Casimir force of thermally excited carriers in a conducting test body that  is an insulator at zero temperature. We thus formulate the following {\it insulator-state prescription}  (ISP) \cite{book2,RMP}:\\

\noindent
{\it insulator-state prescription (ISP): if a  conductor is  in the insulator state (i.e. is an insulator at $T=0$)   free charge carriers should be neglected in the computation of the Casimir force.} \\

This prescription applies  to P-doped Si, if the carrier concentration $n$ is less than the critical density  $n_{\rm cr}$ for the Mott-Anderson insulator-metal transition. Thus, according to this prescription the  contribution of free charges  in the permittivity of conductive Si, represented by the Drude term in  Eq. (\ref{sicon}),  has to be omitted for $n < n_{\rm cr}$  leaving us with:
\be
\epsilon_l^{(2)}\vert_{\rm ISP}=\epsilon^{\rm int}_{\rm Si}(i \xi_l)\;,\;\;\;{\rm for} \;n<n_{\rm cr}\;.\label{plnlow}
\ee
This prescription implies that for $n< n_{\rm cr}$  there is no difference among the permittivities $\epsilon_l$ of the highly resistive Si left sector of the plate and of its right conductive Si sector:
\be
\epsilon_l^{(2)}\vert_{\rm ISP}=\epsilon_l^{(4)}\;,\;\;\;\;\;{\rm for} \;n<n_{\rm cr}\;.
\ee 
This relation  implies at once that within the insulator-state prescription the force difference $F_{\rm diff}$ in our setup vanishes for $n<n_{\rm cr}$:
\be
F_{\rm diff}\vert_{\rm ISP}=0\;,\;\;\;{\rm for}\; n < n_{\rm cr}\;.
\ee
We thus see that the insulator-state prescription leads to a sharp prediction: for $n<n_{\rm cr}$ the measured force difference $F_{\rm diff}$  is zero!
 
\section{Numerical computations}

In  this Section we present the results of our numerical computations of the force $F_{\rm diff}(a)$ for the setup of Fig. \ref{setup}. We have considered three different values for the carrier density of the conductive P-doped Si, i.e. $n=5 \,n_{\rm cr}$, $n=2 \,n_{\rm cr}$ and $n=0.5 \,n_{\rm cr}$ where $n_{\rm cr}=3.84 \times 10^{18} {\rm cm}^{-3}$ is, we recall, the critical density for the Mott-Anderson insulator-metal transition. We note that for the above values of the carrier density,  the Debye radius $R_D=\sqrt{\overline{\epsilon} k_B T/4 \pi e^2 n}$, where $\overline{\epsilon}=\epsilon^{\rm int}_{\rm Si}(0)$ is the bare dielectric constant (i.e. not including the carriers contribution) is always much smaller than the separations we shall consider. For example, for the smallest considered density  $n=0.5 \,n_{\rm cr}$  the Debye radius is  $R_D= 3$ nm,  while the minimum separation that we consider is $a = 100$ nm. Since in all cases $R_D \ll a$  the influence of spatial dispersion  can be safely neglected when considering the contribution of charge carriers to the material response of Si  \cite{pita}, and the local form of Lifshitz theory based on the standard Fresnel reflection coefficients Eqs. (\ref{fresnel})  is fully adequate.

\begin{figure}
\includegraphics [width=.9\columnwidth]{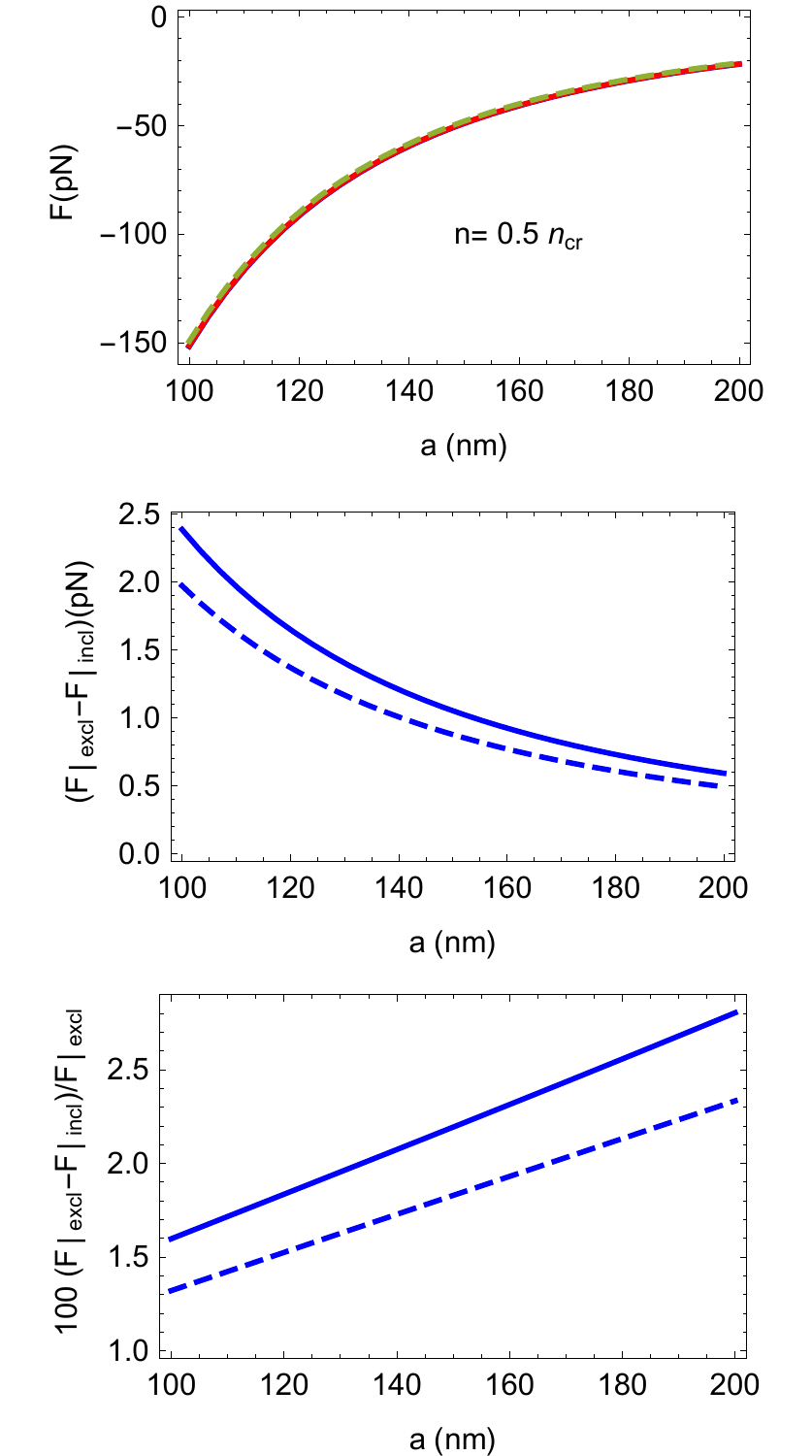}
\caption{\label{Absfor}  Plots of the room temperature Casimir force (in pN) between a gold coated sphere of radius $R=150\,\mu$m and a thick slab of conductive P-doped Si, with carrier concentration $n=0.5\,n_{\rm cr}$.  The blue,  red (light gray) and dashed lines in panel 1 correspond to the plasma, Drude and ISP prescription respectively. The second panel displays the difference $F|_{\rm excl}-F|_{\rm incl}$ among the forces which result by including or excluding the contribution of free carriers. The force $F|_{\rm incl}$ is computed using either the plasma prescription (solid line) or the Drude model (dashed line). The bottom panel shows the magnitude of $F|_{\rm excl}-F|_{\rm incl}$ in percent of the absolute force.}
\end{figure}

In Fig. \ref{Absfor} we show a plot of the room-temperature Casimir force between a Au sphere and a thick slab of P-doped conductive Si, with a carrier concentration $n=0.5\,n_c$ The blue, red and dashed lines in panel 1 correspond to the plasma, Drude and ISP prescription respectively. The second panel displays the difference $F|_{\rm excl}-F|_{\rm incl}$ among the forces which result by including or excluding the contribution of free carriers. The force $F|_{\rm incl}$ is computed using either the plasma prescription (solid line) or the Drude prescription (dashed line). The bottom panel shows the magnitude of $F|_{\rm excl}-F|_{\rm incl}$ in percent of the absolute force. The Figure shows that the differences among the forces predicted by the three prescriptions differ by less than 2 pN in the separation region from 100 to 200 nm, representing a change by less than 2.5 percent in the magnitude of the force. It is clear that a discrimination among the three prescriptions based on an absolute force measurement is extremely difficult. Below we show that the differential setup proposed in this work   engenders a large amplification of the difference among the three prescriptions. 

\begin{figure}
\includegraphics [width=.9\columnwidth]{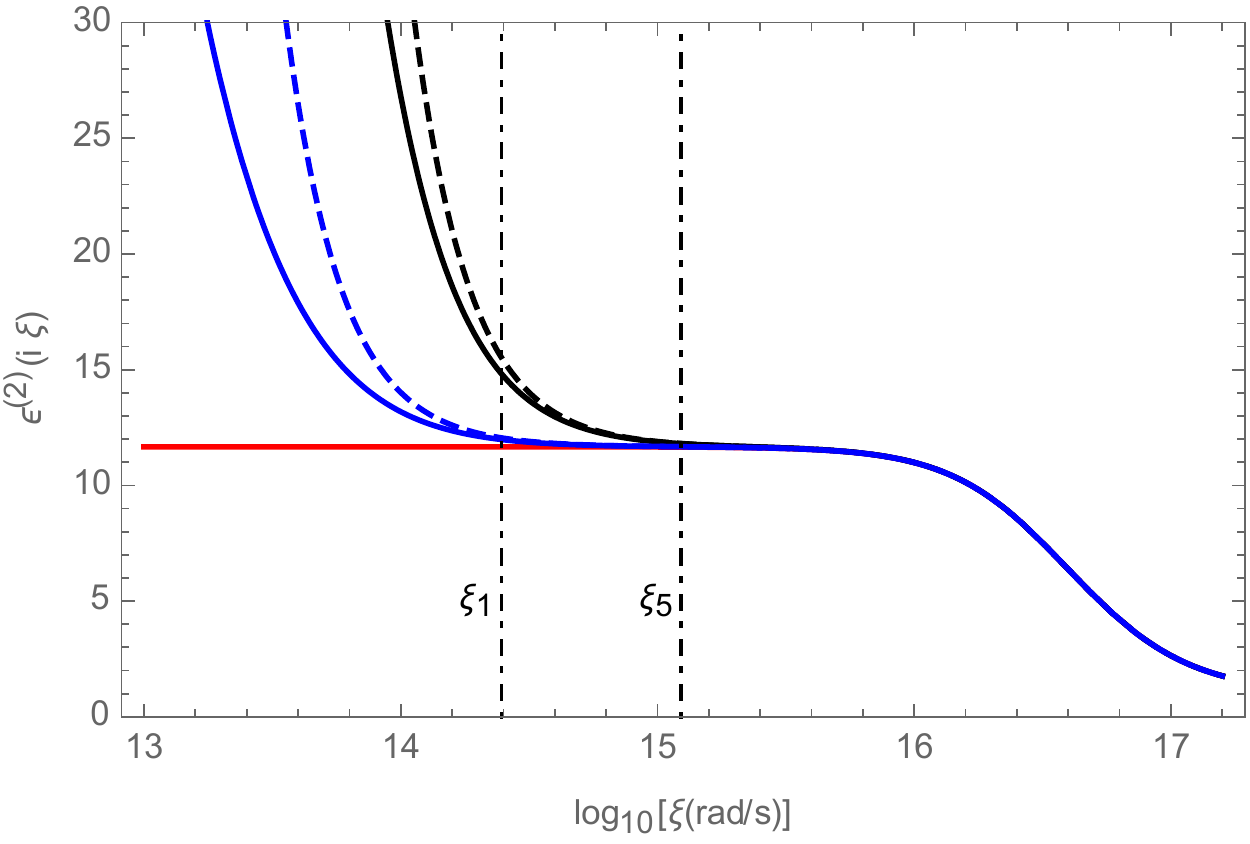}
\caption{\label{epsilon}  Plots of the imaginary-frequency permittivity of Si. The three solid lines from bottom to top correspond, respectively,  to intrinsic Si (red line) and to   conductive P-doped Si, with concentrations $n= 0.5\,n_{\rm cr}$ (blue line)  and $n=5\,n_{\rm cr}$ (balck line), where $n_{\rm cr}$ is the critical carrier density for the Mott-Anderson insulator-metal transition. The dashed  lines correspond to neglect of dissipation in the contribution of free carriers (plasma model prescription).}
\end{figure}

Before we turn to the computation of the differential force, it is useful to make the following observation. According to Lifshitz formula,  the Casimir force $F_{\rm diff}(a)$ is expressed by a sum over the Matsubara frequencies $\xi_l=2 \pi l k_B T/\hbar$. The number of Matsubara frequencies that contribute significantly to the (absolute) Casimir force between two dielectric bodies at distance $a$ in vacuum, can be estimated to be around 10 $\omega_c/\xi_1$, where $\omega_c=c/a$ is the characteristic frequency of the system. For the minimum separation $a=100$ nm that we are going to consider, this corresponds to something like 120 terms.  It is easy to see that far less Matsubara frequencies  contribute significantly to the force difference $F_{\rm diff}$ in our setup. This can be understood by looking at Fig. \ref{epsilon},  which shows plots of the  permittivities along the imaginary axis of intrinsic Si (the red curve)  and of conductive P-doped Si for two of the three values of the concentration that we considered, i.e. for $n=5\,n_{\rm cr}$ (black solid line) and for $n=0.5\,n_{\rm cr}$ (blue solid line). The  dashed blue and black curves  correspond to neglecting in Eq. (\ref{sicon})  the relaxation frequency $\gamma_2$ in the Drude term, and so they represent the permittivities that are used to compute the force within the plasma prescription. The dot-dashed vertical left and right lines shown in the Figure correspond, respectively, to the first and to the fifth Matsubara mode, i.e. to $\xi_1$ and $\xi_5$.
The Figure clearly shows that the five permittivities are practically undistinguishable for frequencies $\xi > \xi_5$, and this implies that only the  first five or so Matsubara terms, contribute significantly to the force difference $F_{\rm diff}$. This is very good news for us, because the effect we are after has a  {\it low-frequency} character, and thus the fact that the differential measurement is insensitive to the uninteresting high-frequency region of the spectrum represents a big plus for the proposed setup.  This feature of the apparatus makes it unnecessary to have detailed information on the optical properties of the materials, whose incomplete or inaccurate knowledge represents a source of theoretical uncertainty in absolute force measurements. 

\begin{figure}
\includegraphics [width=.9\columnwidth]{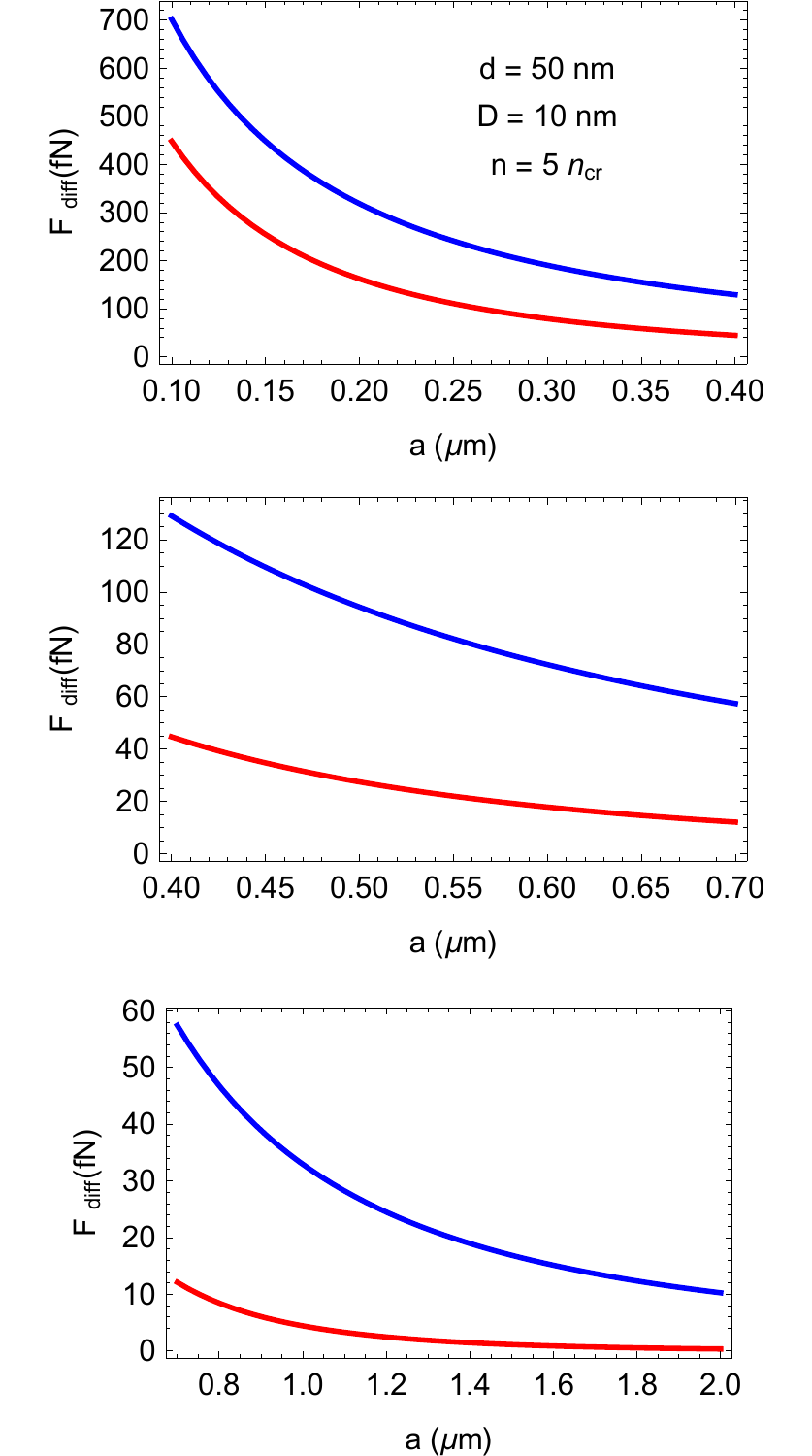}
\caption{\label{force1}  Plots of the room temperature ($T=300$ K) force $F_{\rm diff}$ (in fN)  versus separation $a$, for  carrier density $n= 5\,n_{\rm cr}$. The three panels show $F_{\rm diff}$ in different separation ranges. In all panels, the lower red curves correspond to the Drude model, while  the upper blue curves correspond to the  plasma prescription.}
\end{figure}

\begin{figure}
\includegraphics [width=.9\columnwidth]{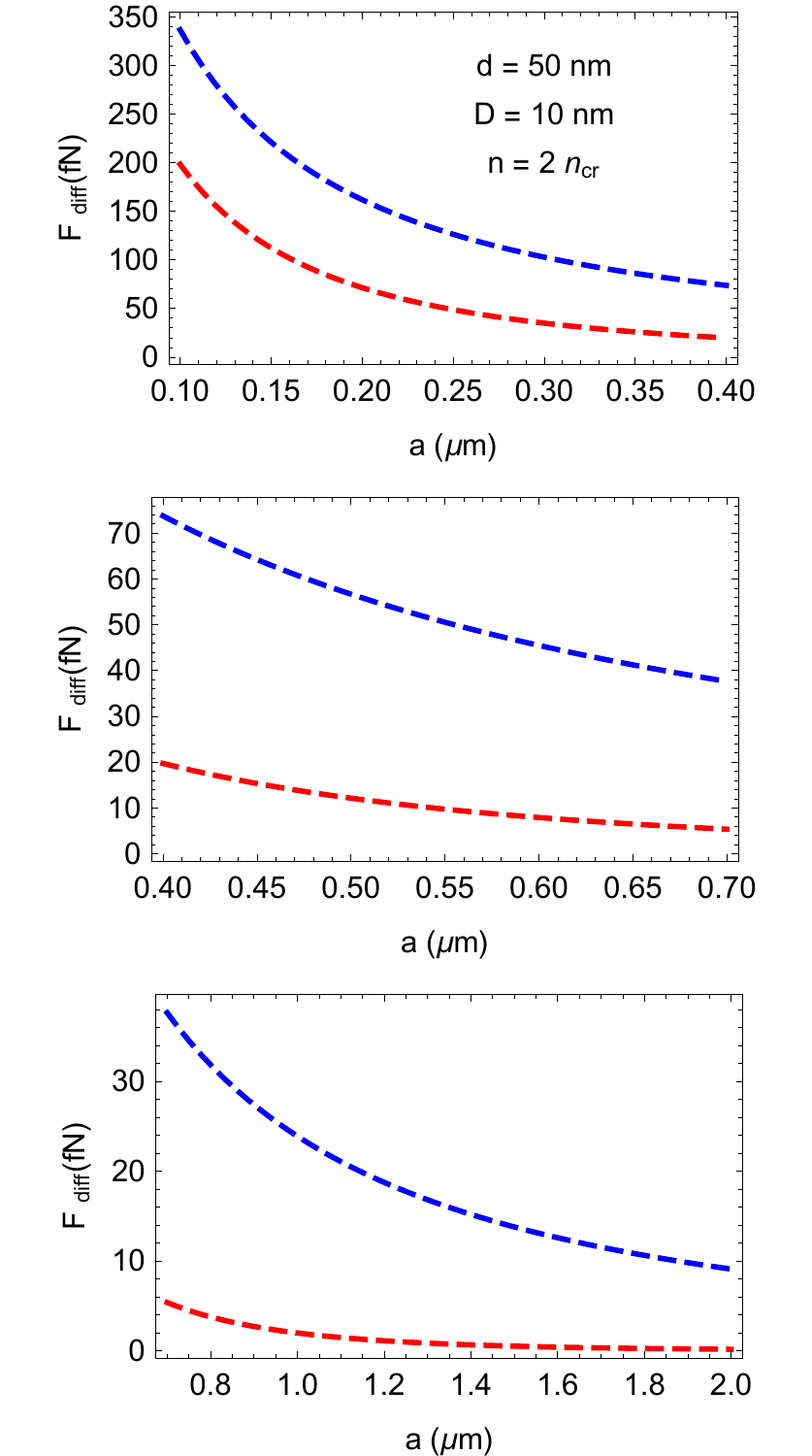}
\caption{\label{force2}  Plots of the room temperature ($T=300$ K) force $F_{\rm diff}$ (in fN)  versus separation $a$, for  carrier density $n= 2\,n_{\rm cr}$.  The three panels show $F_{\rm diff}$ in different separation ranges. In all panels, the lower red curves correspond to the Drude model, while the upper blue curves correspond to the   plasma prescription.}
\end{figure}

\begin{figure}
\includegraphics [width=.9\columnwidth]{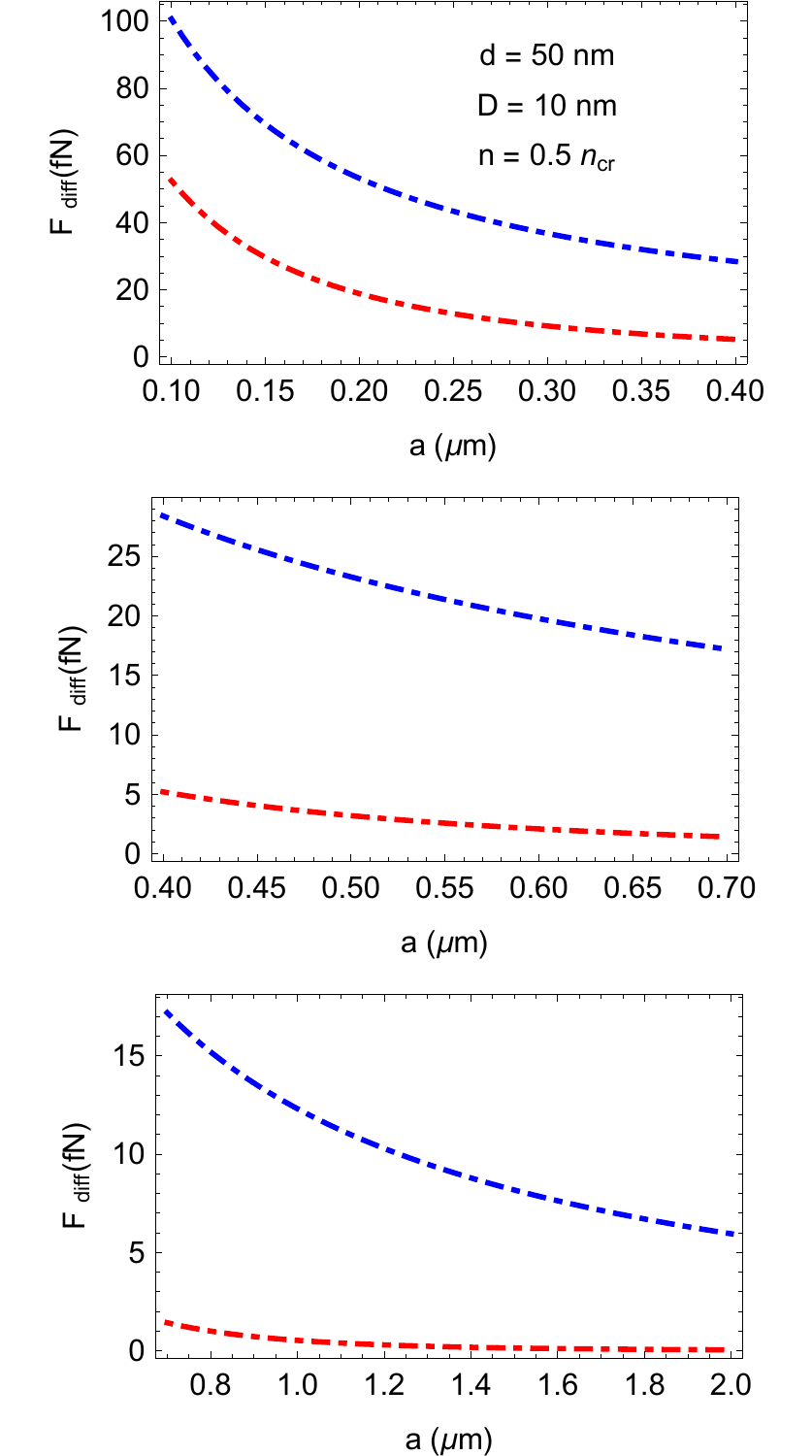}
\caption{\label{force3}  Plots of the room temperature ($T=300$ K) force $F_{\rm diff}$ (in fN)  versus separation $a$, for  carrier density $n= 0.5\,n_{\rm cr}$.  The three panels show $F_{\rm diff}$ in different  separation ranges. In all panels, the lower red curves correspond to the Drude model, while the upper blue curves correspond to the   plasma prescription. The insulator state prescription predicts a null force $F_{\rm diff}=0$.}
\end{figure}

We computed the force $F_{\rm diff}(a)$ for room temperature ($T=300$ K) in the separation range $100 \,{\rm nm} < a < 2\,\mu{\rm m}$.  Plots of the force $F_{\rm diff}$ (in fN) versus separation are shown in Fig.  \ref{force1}  for $n= 5\, n_{\rm cr}$, in  Fig.  \ref{force2}  for $n= 2\, n_{\rm cr}$ and in  Fig.  \ref{force3}  for $n= 0.5 \,n_{\rm cr}$.
In all panels of  these three figures, the lower red curves and the upper blue curves  correspond  to the Drude and plasma prescriptions, respectively. We remark that for $n=0.5\,n_{\rm cr}$ the ISP prescription predicts a null force $F_{\rm diff}\vert_{\rm ISP}=0$. These figures show that the three  prescriptions   lead to widely different predictions  for the differential force $F_{\rm diff}$, that should be easily distinguishable in a wide range of separations with an apparatus having a fN sensitivity.

\begin{figure}
\includegraphics [width=.9\columnwidth]{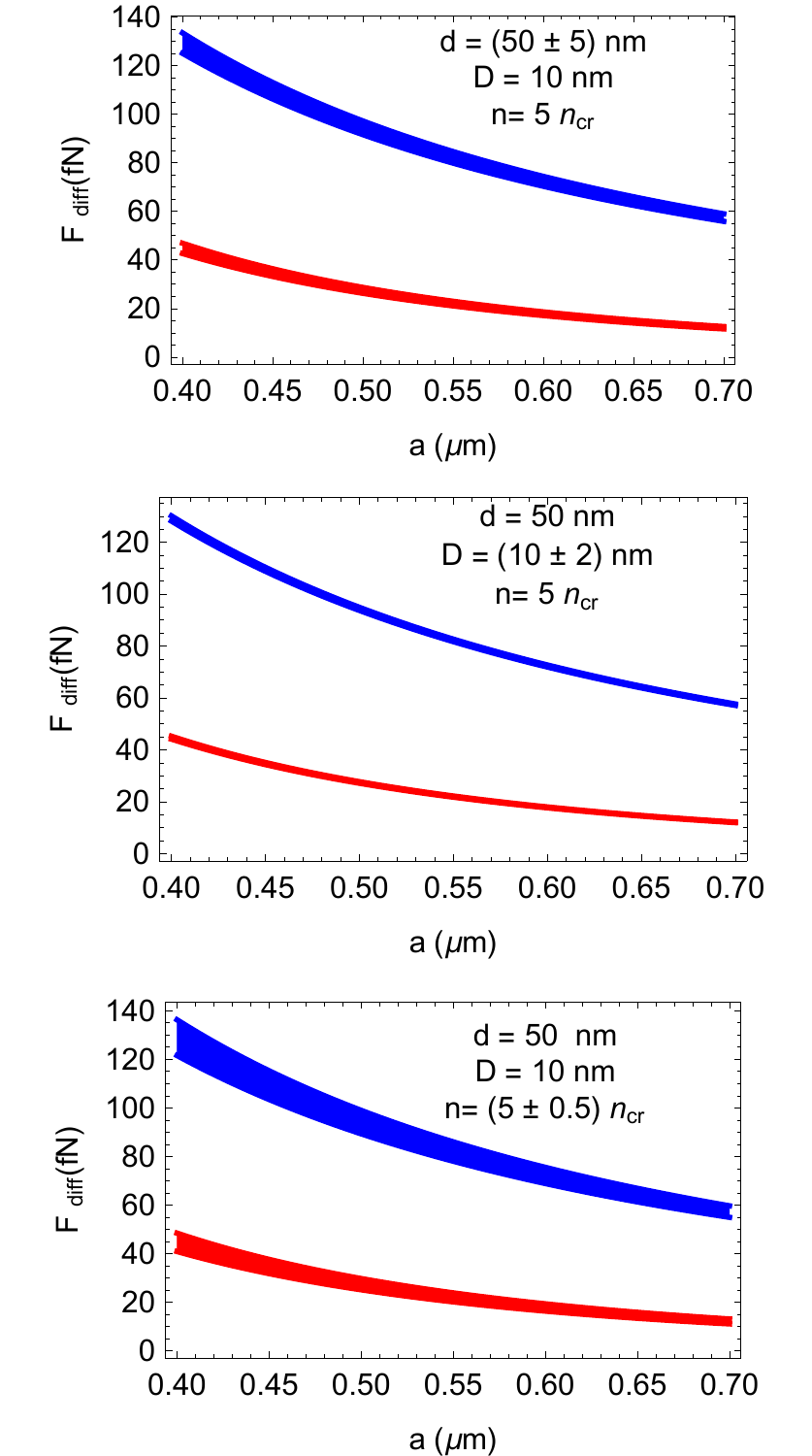}
\caption{\label{band1}  Bands of variation of the differential force for $n=5\,n_{\rm cr}$, corresponding to a ten percent uncertainty in the thicknesses $d$   conductive Si over-layer (upper panel) in the thickness of the SiO$_2$ layer (middle panel) and in the carrier density $n$ (lower panel). In all panels, the lower red bands correspond to the Drude model, while  the upper blue bands correspond to the  plasma prescription.}
\end{figure}

\begin{figure}
\includegraphics [width=.9\columnwidth]{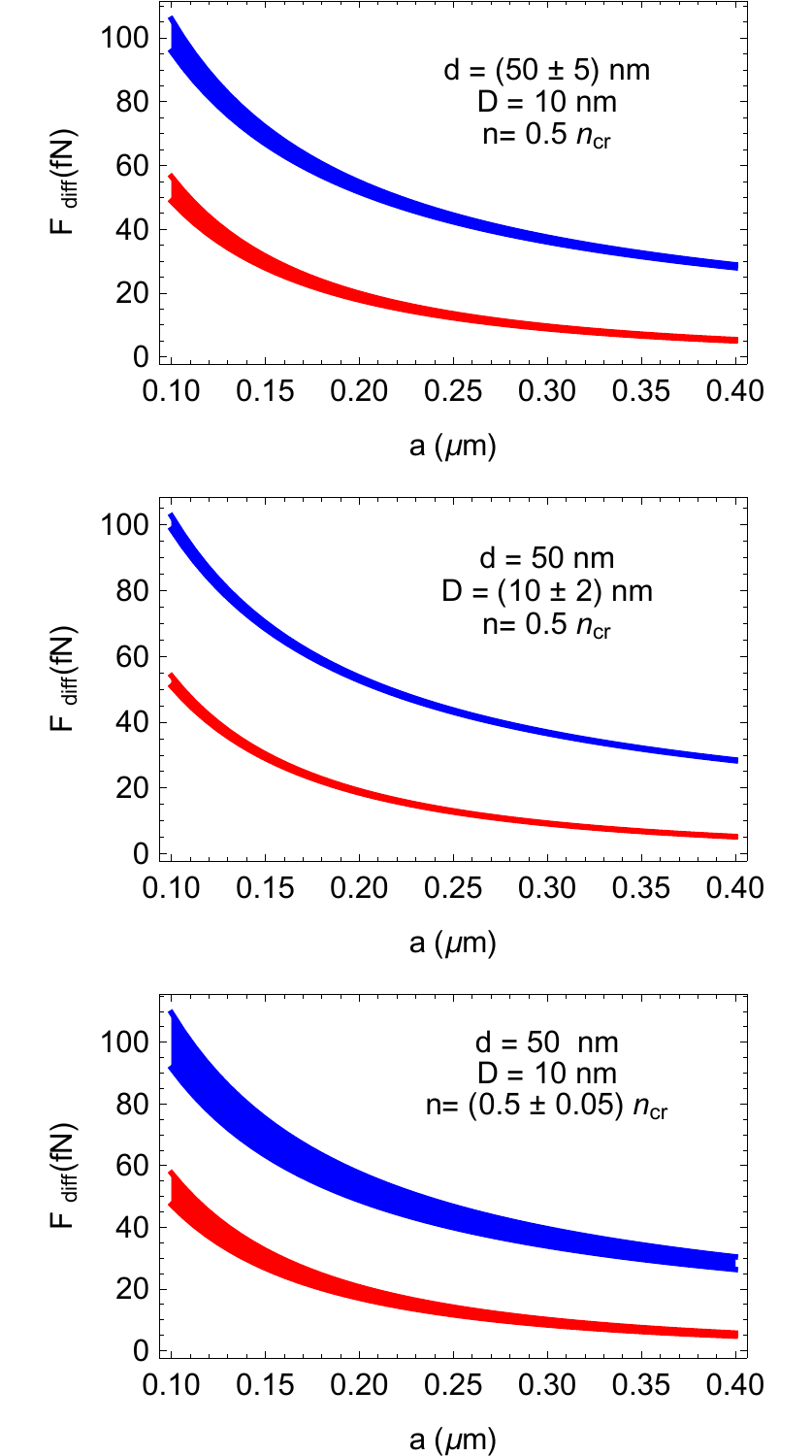}
\caption{\label{band2} Bands of variation of the differential force for $n=0.5\,n_{\rm cr}$, corresponding to a ten percent uncertainty in the thicknesses $d$   conductive Si over-layer (upper panel) in the thickness of the SiO$_2$ layer (middle panel) and in the carrier density $n$ (lower panel).    In all panels, the lower red bands correspond to the Drude model, while  the upper blue bands correspond to the  plasma prescription.   No band of variation is displayed for the insulator state prescription, since this prescription predicts a null differential force $F_{\rm diff}=0$, irrespective of the thicknesses $d$ and $D$, and of the carrier density $n$ (provided that $n$ remains less than $n_{\rm cr}$).}
\end{figure}

We have checked that the predicted differential force is robust against systematic  errors arising from uncertainties in both geometric and material-dependent parameters that characterize our setup. This is demonstrated by Fig. \ref{band1} and Fig. \ref{band2}, which show the bands of variation of the differential force, corresponding to a ten percent uncertainty in the thickness $d$ of the  conductive Si over-layer (upper panels), in the thickness of the SiO$_2$ layer (middle panels), and in the carrier density $n$ (lower panels). In all panels, the lower red bands and the upper blue bands correspond, respectively, to the Drude and plasma prescriptions. Notice that in Fig. \ref{band2} no band of variation is displayed for the ISP, since this prescription predicts a null differential force $F_{\rm diff}=0$, irrespective of the thickness $d$ and $D$, and on the carrier density $n$ (provided that $n$ remains less than $n_{\rm cr}$). The displayed graphs show that the parameter that needs to be better controlled is the concentration of dopant $n$.

\begin{figure}
\includegraphics [width=.9\columnwidth]{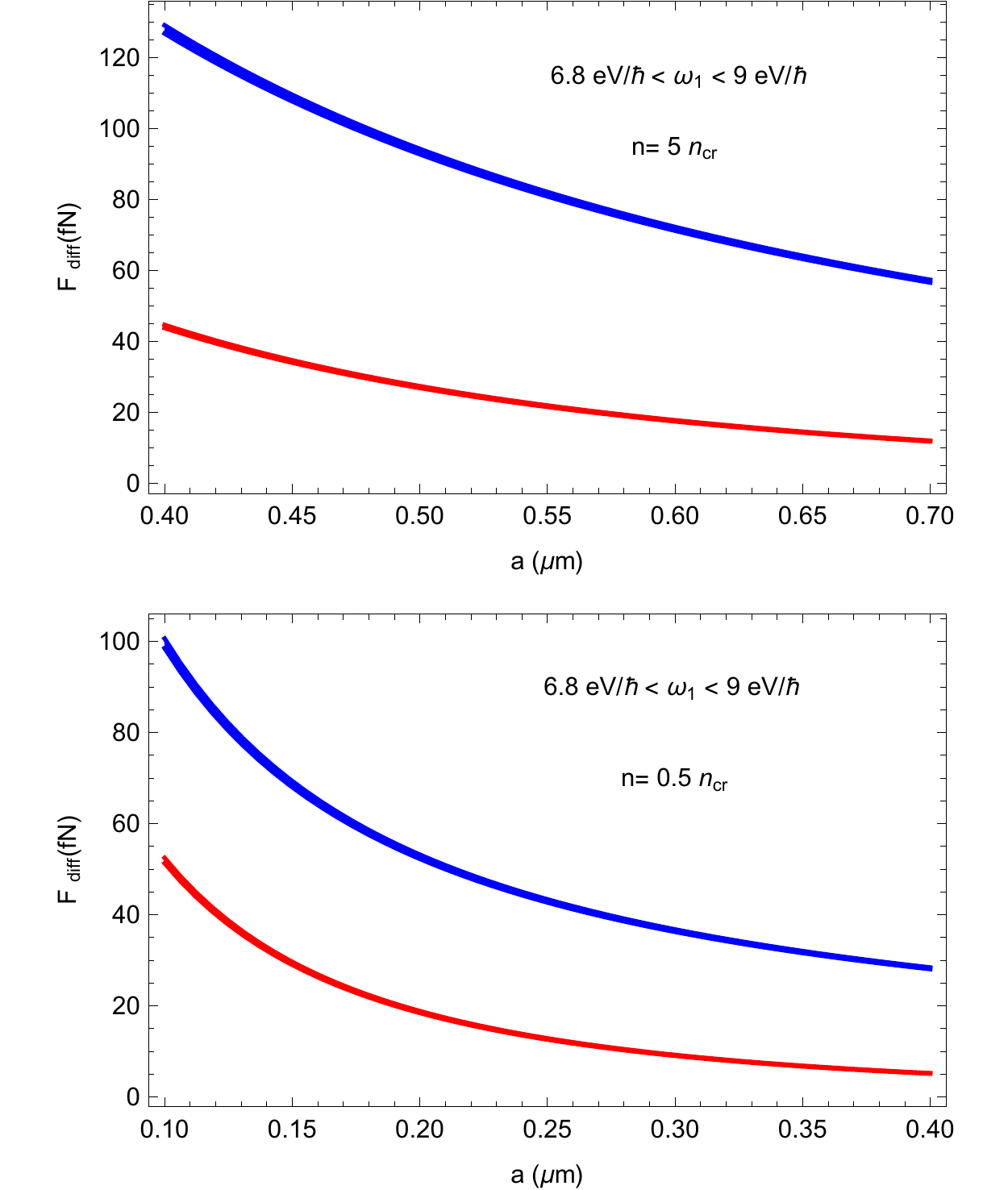}
\caption{\label{band3} Bands of variation of the differential force corresponding to values of the Au plasma frequency in the interval 6.8 eV/$\hbar < \omega_1 <$ 9 eV/$\hbar$.    In all panels, the lower red bands correspond to the Drude model, while  the upper blue bands correspond to the  plasma prescription.  For the carrier concentration $n= 0.5\,n_{\rm cr}$  no band of variation is displayed for the insulator state prescription, since this prescription predicts a null differential force $F_{\rm diff}=0$, irrespective of the optical properties of Au.}
\end{figure}

Finally, in Fig. \ref{band3} we show the band of variation of the differential force  corresponding to an uncertainty in the plasma frequency $\omega_1$ of the gold sphere. The sample variation of the Au plasma  frequency  has been  much debated in the literature (most Casimir experiments use Au test bodies), for it has been shown that an inaccurate determination of this parameter  may by itself lead to  a large theoretical error, as large as five percent,  on the magnitude of the Casimir force \cite{vitaly}. In order to reduce this source error modified dispersion relations have been devised \cite{bimonteK,bimonteKK} that  
suppress the influence of low frequencies on the determination of the permittivity for imaginary frequencies.
It is fortunate that this problem is irrelevant to the present scheme, for the differential force is  weakly dependent on the value of $\omega_1$. The narrow bands shown in Fig. \ref{band3} (in all panels, the lower red bands and the upper blue bands correspond to the Drude and plasma prescriptions respectively) correspond to the wide interval 6.8 eV/$\hbar < \omega_1 <$ 9 eV/$\hbar$ which includes all sample-dependent values of the plasma frequency that have been reported in the literature \cite{vitaly}. No band is shown in Fig. \ref{band3} for $n=0.5\,n_{\rm cr}$ because the differential force is zero within the ISP, independently of the properties of the Au sphere. 

\section{Conclusions}

Over the last 20 years, intense experimental and theoretical investigations of the Casimir effect with conducting test bodies raised puzzling questions about the influence of free charge carriers on the strength of the Casimir force. Theoretical predictions based on Lifshitz theory of dispersion forces between dielectric test bodies appear to be in disagreement with the most precise experiments  \cite{decca1,decca2,decca3,decca4,chang,bani1,bani2}.  It appears that  in order to bring experimental data into agreement with Lifshitz theory, one has to abandon the natural prescription based on the fluctuation-dissipation theorem of statistical physics,  to account for the effect of conductance on the Casimir force. Agreement with data can be achieved by neglecting the effect of relaxation on the free carriers into Lifshitz formula, which means that as far as the Casimir effect is concerned free charges in conducting test bodies behave as a dissipation-less plasma. 

Semiconductors offer a unique opportunity to investigate the influence of conduction on the Casimir effect \cite{umar2006, umar2006bis,umar1}, since their conductivity can be modified by many orders of magnitude by doping. It has been known for a long time \cite{rosen} that doped semiconductors undergo a Mott-Anderson metal-insulator transition, when the concentration of dopant atoms exceeds a critical density $n_{\rm cr}$. It is of great interest to investigate whether the metal-insulator transition has any bearing onto the Casimir effect. The answer to this question depends crucially on the prescription that is used to include the effect of free carriers in doped semiconductors on the Casimir force. According to the standard prescription, based on the fluctuation-dissipation theorem, no effect is to be expected since the optical properties of semiconductors at room temperature  do not change appreciably across the transition.  A  different prescription, based on a thermodynamic  argument \cite{vladimir},  suggests that free carriers contribute to the Casimir force when the semiconductor is in the metallic state, i.e. for doping levels higher than the critical one, while they should be excluded for concentrations less than the critical one. This prescription implies that for the Casimir force a discontinuous   change  across the metal-insulator transition!  

It is clearly of great interest to see if this bold prediction can be tested experimentally. Observation of the effect by conventional  Casimir apparatus, based on absolute  force measurements, is very hard because the effect is predicted to be  small, perhaps one or two percent in the submicron separation range where Casimir experiments are most precise.

In this paper we have described an {\it isoelectronic differential apparatus} that should allow for an easy observation of the effect. The crucial ingredient of the setup is a micro-fabricated patterned Si plate, whose left half is made of highly resistive Si, while its right half is made of P-doped Si. A key feature of the patterned plate is the presence of a P-doped thin Si over-layer of uniform thickness, that covers both halves of the plate. The purpose of the over-layer is to screen out possible inhomogeneities in the potential patches that may exist on the surface of the left and right Si slabs, as a result of their different doping levels. 
The proposed experiment consists in a differential measurement of the force experienced by a Au coated sphere, as it is moved from the undoped left half to the doped right half of the Si plate. The differential character of the measurement ensures automatic cancellation of several effects that plague ordinary Casimir setups, like errors in the sphere-plate separation, roughness, and potential patches. We have also checked that the apparatus is robust against possible uncertainties in the parameters that characterize it, like the thickness of the over-layer, the concentration of dopants, and errors in the optical properties of the materials. In this work we considered for brevity only the case of P-doped Si. Doping the right section of the Si plate by other elements like sulfur \cite{vladimir}, might lead to a larger differential force and/or a better discrimination among the three theoretical prescriptions for computing the Casimir force. We leave the optimization of the setup for a future work. 

In view of  the fN sensitivity reached by current differential Casimir apparatus  \cite{ricardomag,umardiff}, the numerical computations presented in this work show that the proposed scheme should allow for a clear discrimination among  alternative theories for the Casimir effect in doped semiconductors.  Observation of the effects described in this paper would  shed much light on the puzzling and yet unresolved problem of the influence of conductivity on the Casimir effect.

\acknowledgments

The author thanks G. L. Klimchitskaya and V. M. Mostepanenko  for enlightening correspondence during the manuscript preparation, and T. Emig, R. L. Jaffe, N. Graham, M. Kardar and M.  Kr\"uger for discussions.  

 .

\end{document}